\documentclass[11pt]{article}
\usepackage{amsfonts,amsgen,amsmath,amssymb,amsbsy,geometry,graphicx,enumerate,psfrag,subfigure}

\newcommand{\got}[1]{\mathcal{#1}}

\newcommand{\pd}{\partial}

\newcommand{\vc}[1]{{\boldsymbol{#1}}}

\newcommand{\fbrl}{\Biggl\{}
\newcommand{\fbrr}{\Biggr\}}
\newcommand{\sbrl}{\Biggl[}
\newcommand{\sbrr}{\Biggr]}

\newcommand{\SSC}{Solid State Commun.}
\newcommand{\PR}{Phys. Rev.}
\newcommand{\MST}{Meas. Sci. Technol.}
\newcommand{\NJP}{New J. Phys.}
\newcommand{\JASA}{J. Acoust. Soc. Am.}
\newcommand{\APL}{Appl. Phys. Lett.}
\newcommand{\PRS}{Proc. R. Soc.}
\newcommand{\JEWA}{J. of Electromagn. Waves and Appl.}
\newcommand{\OC}{Opt. Commun.}
\newcommand{\ZAMM}{Z. Angew. Math. Mech.}

\newcounter{mylistcounter1}

\newcounter{mylistcounter2}
\newenvironment{mylist2}{\begin{list}{(\alph{mylistcounter2})}
	{\usecounter{mylistcounter2}\setlength\labelwidth{3cm}\setlength\leftmargin{2.35cm}\setlength\itemsep{5pt}}}
{\end{list}}

\title{Scattering by coupled resonating elements in air}

\begin{document}

\author{%
  Anton Krynkin$^1$, Olga Umnova$^1$,\\ Alvin Y.B. Chong$^2$, Shahram Taherzadeh$^2$ and Keith Attenborough$^2$\\
	{\small $^1$ Acoustics Research Centre, The University of Salford, Salford, Greater Manchester, UK}\\
	{\small $^2$ Department of Design Development Environment and Materials, The Open University, Milton Keynes, UK}\\
	{\small email: a.krynkin@salford.ac.uk}
}

\maketitle

\begin{abstract}

Scattering by (a) a single composite scatterer consisting of a concentric arrangement of an outer N-slit rigid cylinder and an inner cylinder which is either rigid or in the form of a thin elastic shell and (b) by a finite periodic array of these scatterers in air has been investigated analytically and through laboratory experiments. The composite scatterer forms a system of coupled resonators and gives rise to multiple low frequency resonances. The corresponding analytical model employs polar angle dependent boundary conditions on the surface of the N-slit cylinder. The solution inside the slits assumes plane waves. It is shown also that in the low-frequency range the N-slit rigid cylinder can be replaced by an equivalent fluid layer. Further approximations suggest a simple square root dependence of the resonant frequencies on the number of slits and this is confirmed by data. The observed resonant phenomena are associated with Helmholtz-like behaviour of the resonator for which the radius and width of the openings are much smaller than the wavelength. The problem of scattering by a finite periodic array of such coupled resonators in air is solved using multiple scattering techniques. The resulting model predicts band-gap effects resulting from the resonances of the individual composite scatterers below the first Bragg frequency . Predictions and data confirm that use of coupled resonators results in substantial insertion loss peaks related to the resonances within the concentric configuration. In addition, for both scattering problems experimental data, predictions of the analytical approach and predictions of the equivalent fluid layer approximations are compared in the low-frequency interval.

\end{abstract}




\section{Introduction}

Infinite periodic structures do not support wave propagation in certain frequency intervals known as stop bands (band gaps) \cite{Sigalas, Nicorovici}. For finite arrays of periodically arranged scatterers, referred to as Sonic Crystals, the stop bands correspond to the frequency intervals of very low transmission \cite{Martinez-Sala, Miyashita, Umnova}. The position of the stop bands can be tuned to selected frequency intervals by changing the spacing between the scatterers. This makes Sonic Crystals attractive for applications as noise barriers particularly for narrow band sources.

The performance of the Sonic Crystals can be improved by increasing the filling fraction \cite{Nicorovici, Caballero}, varying the arrangement of the scatterers \cite{Romero-Garcia} and also by replacing the scatterers with the resonant elements \cite{Liu, Movchan_Guenneau, XHu, Sainidou, Kosevich, Guenneau, Fuster-Garcia, Cui, SUOU}. The latter allows increasing insertion loss in the low-frequency range if the resonant frequencies of the scatterers lie below the first stop band associated with the array periodicity. The most obvious choice for such resonant elements are split ring resonators \cite{Movchan_Guenneau, Guenneau, Cui} which essentially are 2D Helmholtz resonators. Alternatively an array of thin elastic shells in air has been shown to possess strong multiple resonances in the low-frequency range \cite{Sainidou, SUOU}.

Arrays of composite scatterers such as concentric split ring resonators \cite{Movchan_Guenneau, Smith} are shown to support band gaps generated by the multiple resonances of each scatterer in the array. In this paper we investigate the concentric arrangement of two types of the resonant elements i.e. thin elastic shells and split ring resonators with multiple slits, referred to as N-slit cylinders. These composite scatterers, referred to as composites, serve a double purpose. First, they support multiple resonances including Helmholtz and annular cavity  resonances and also axisymmetric resonance of the elastic shells. The latter is defined by the geometry of the shell and not by the area of the cavity that therefore gives advantage in tuning of the composite multiple resonant behaviour compared to that for the concentric split ring resonators. It is also shown that the coupling between two resonators leads to the shift of the axisymmetric shell resonance to the lower frequency range. The analogous effect for coupled mechanical oscillator (multiple degree of freedom system) is well known \cite{Morse}. Second, for the purpose of creating a noise barrier an array of elastic shells exposed to the environment would not be practical. A simple solution is to protect the shells by using outer cylinders containing slits surrounding inner elastic shells arranged concentrically.

The scattering problem for the hollow cylinder with non-uniform boundaries has been solved semi-analytically in \cite{Tsalamengas,Yin} and references therein. This involves solving the linear algebraic system of equations derived from the analytical evaluation of the integral equations. An alternative approach has been developed in grating theory \cite{Montiel,Guizal} where solution of the studied partial differential equations is subject to the boundary conditions dependent on the coordinates.

Three configurations involving the N-slit cylinder are studied in this paper (a) alone, (b) concentric with an inner rigid cylinder and (c) concentric with an inner elastic shell. To solve the scattering problem for N-slit cylinders we use boundary conditions dependent on polar angle \cite{Montiel}. In section \ref{ss} this approach is employed to solve a single composite scatterer problem that is similar to that solved for an electromagnetic case \cite{Guizal}. However in the present model the finite thickness of the wall of the N-slit cylinder is taken into account. This avoids the use of adjustable parameters introduced for numerical stability \cite{Montiel,Guizal} and provides a more accurate description of the real structure used in the experiments reported later. The solution inside slits is replaced by jump conditions \cite{Rawlins} that describe the slit interface as a moving piston. It should be noted that use of jump conditions to replace the slits makes it possible to accurately predict the total wave field in the low- and mid-frequency range.

It is also shown that in the low-frequency range the slit cylinder can be replaced by an equivalent fluid layer. The approximations are based on the results derived for the perforated plate \cite{Lamb, Smits, Shenderov}. The developed approximations are of particular interest for studying the low-frequency resonant behaviour of the single composite arrangement and a finite periodic array of them. This behaviour is attributed to the Helmholtz-like resonances with wavelength much bigger than the characteristic sizes of the cylinder (i.e. radius and width of the slits).

Experimental validation of the analytical and numerical predictions has been carried out. In particular the results obtained with the grating theory are compared with the low-frequency approximations for the cylinders with different number of slits and various inner structure (N-slit cylinder alone, rigid core and elastic shell core). However the analytical models could be used also to study the acoustical properties of arrays incorporating resonators of various thickness and with different slit width.

In section \ref{aos} the model is generalised for finite arrays of N-slit cylinders and composite elements. In section \ref{experiment} the experimental setup is described and the model predictions are compared with the data.

\section{Single scatterers}\label{ss}

\subsection{Formulation}\label{ssformulation}

\begin{figure}[t]
		\center
		\subfigure[]{\includegraphics[scale=.7]{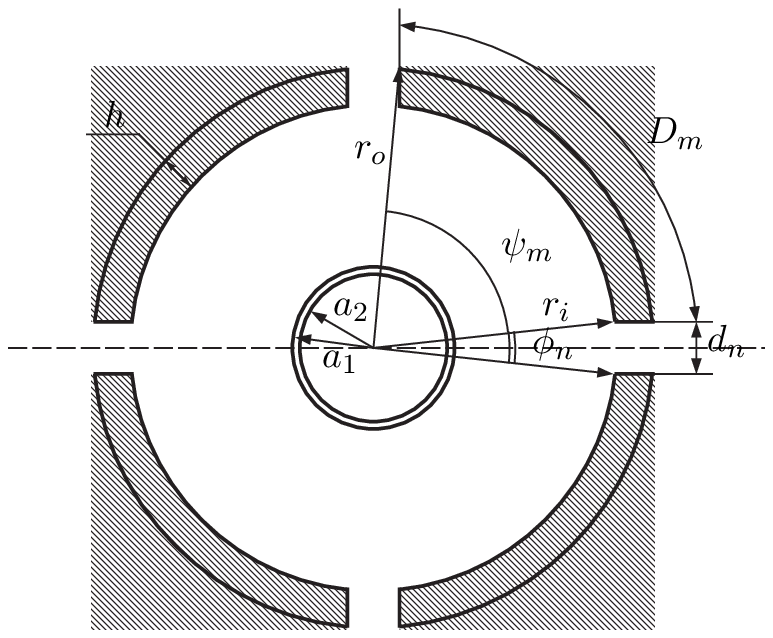}}\hspace{5mm}
		\subfigure[]{\includegraphics[scale=.7]{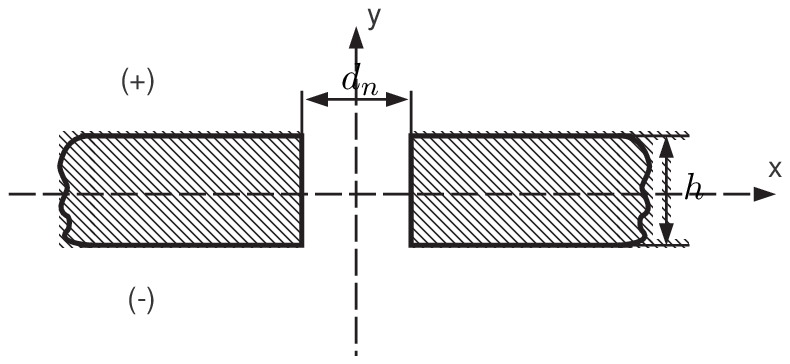}}
    \caption{(a) Cross-section of composite element consisting of a concentric arrangement of an outer 4-slit rigid cylinder and an inner elastic cylindrical shell. (b) Geometry of the slit.}
    \label{fig:comp_geom}
\end{figure}

Consider the two-dimensional problem of acoustic wave scattering by a single N-slit rigid cylindrical shell of thickness $h$ and external radius $r_{o}$. The sound is generated by the cylindrical point source placed at the origin of the coordinate system which is defined by either the Cartesian $(x,y)$ or polar $(r,\theta)$ coordinates. Throughout the paper the time-harmonic dependence is taken as $\exp(-i \omega t)$. The widths of consecutive slits in $Oxy$ plane are denoted by $d_n,\,n=1..N$ and they are infinitely long in the direction of the cylinder main axis $Oz$, see Figure~\ref{fig:comp_geom}(a). The external length of the rigid arc is given by $D_n$. It is also assumed that the thickness of the rigid shell is much smaller than its radius so that the following geometric simplifications can be applied:
\begin{mylist2}
	\item Internal and external arc lengths of the rigid strip have the same length;
	\item Internal and external arcs subtend the same angle.
	\item The angle $\phi_n$ subtended by the arc of n-th slit and its width $d_n$ are related by $d_n=\phi_n r_o,\,n=1..N$
	\item The angle $\psi_m$ subtended by the arc between slit (n-1) and slit n and its width $D_m$ are related by $D_m=\psi_m r_o,\,m=1..N$
\end{mylist2}
Without loss of generality it is assumed that radius of the vector passing through the middle of the first slit makes either zero or $\pi$ angle with the $Ox$ axis.

For simplicity, the acoustic environments outside, inside of the N-slit cylinder and in the slits are assumed to be identical and are described  by density $\rho$ and sound speed $c$. We consider three configurations: (a) an N-slit cylinder alone, (b) with a rigid cylindrical core inside and (c) with an elastic shell inside.

The displacement potential $p(\vc{r})$ in the acoustic medium satisfies the Helmholtz equation
\begin{equation}
\label{ss_Helmholtz}
	\Delta p_\alpha + k^2 p_\alpha=0,
\end{equation}
where $k=\omega/c$ is the wavenumber, index $\alpha$ relates solution $p$ to one of the regions (i.e. '$o$' outside the configuration, '$i$' inside it and '$s$' within the slits) and Laplacian $\Delta$ is given by either $\displaystyle{\frac{1}{r}\frac{\pd}{\pd r}\left(r\frac{\pd}{\pd r}\right)+\frac{1}{r^2}\frac{\pd^2}{\pd \theta^2}}$ or $\displaystyle{\frac{\pd^2}{\pd x^2}+\frac{\pd^2}{\pd y^2}}$.
 
The outer solution $p_o(\vc{r})$ also satisfies Somerfeld's radiation conditions
\begin{equation}
\label{ss_radiation}
	\frac{\pd p_o}{\pd r} - i k p_o = o\left(r^{-1/2}\right),\;\text{as}\; r \rightarrow \infty,
\end{equation}
where $r=\sqrt{x^2+y^2}$.

To proceed with the boundary conditions imposed on the surface of the N-slit cylinder we first solve an auxiliary problem for the slits. The geometry of this problem is illustrated in Figure \ref{fig:comp_geom}(b). Waves propagating inside the slit are described by plane wave solution as 
\begin{equation} 
\label{ss_slitsol}
	p_s=C_1 e^{i k y} + C_2 e^{-i k y},
\end{equation}
where $C_1$ and $C_2$ are unknown coefficients. It is assumed that contribution of the x-dependent components to $p_s$ is negligible when slits are small compared to the wavelength (that is $k d_n \ll 1$ and $k h \ll 1$).


Solution \eqref{ss_slitsol} is subject to the continuity conditions imposed on the slit faces $y=\pm h/2$ as
\begin{align}
\label{ss_slitbc}
	p_\alpha &= p_s, \nonumber\\
	q_\alpha &= \frac{\pd p_s}{\pd y},
\end{align}
where $q_\alpha=\pd p_\alpha/\pd n$ is the normal derivative on the slit faces and index $\alpha=o,i$. Using solution \eqref{ss_slitsol} and its unknown coefficients $C_1$ and $C_2$, equations \eqref{ss_slitbc} can be transformed to the jump conditions \cite{Rawlins}
\begin{align}
\label{ss_slitJC}
	p_i	&= p_o - h q_o,\nonumber\\
	q_i	&= k^2 h p_o + q_o,
\end{align}
that relate the wave field inside the N-slit cylinder to that of the outer region at the slit interface. Note that due to assumption $k h \ll 1$ the trigonometric functions in \eqref{ss_slitJC} are replaced by their leading orders.

The solution of equation \eqref{ss_Helmholtz} is subject to the jump conditions \eqref{ss_slitJC} as well as to the Neumann conditions imposed on the rigid surface of the N-slit cylinder. The former and the latter can be combined into the set of two boundary conditions \cite{Montiel} that are
\begin{align}
\label{ss_BC}
	&\frac{\pd p_i}{\pd r}	= \frac{\pd p_o}{\pd r} + f(\theta) k^2 h p_o,\nonumber\\
	&\frac{\pd p_o}{\pd r}-\frac{f(\theta)}{h}\left(p_o-p_i\right)=0,
\end{align}
where stepwise function $f(\theta)$ of the polar angle $\theta$ introduces the distribution of slits along the N-slit cylinder surface over the interval $\theta\in[0,2\pi]$ as 
\begin{align}
\label{ss_f}
	f(\theta)=\left\{\begin{array}{ll}
							1 & \textrm{if}\quad \theta\in [-\phi_1/2,\phi_1/2]\cup..\cup[2\pi-\phi_1/2-\phi_N-\psi_N,2\pi-\phi_1/2-\psi_N],\\
							0 & \textrm{otherwise.}
						\end{array}\right.
\end{align}
For the non-zero value of this function equations \eqref{ss_BC} are transformed to the jump conditions \eqref{ss_slitJC} whereas the Neumann conditions can be obtained from \eqref{ss_BC} by setting $f(\theta)$ to zero.

The periodic angle distribution of N slits can be introduced through the alternative form of $f(\theta)$ that is
\begin{align}
\label{ss_fprs}
	f(\theta)=&\sum_{n=2}^{N}\left\{ H\left(\theta-\frac{2\pi (n-1)}{N}+\phi_n/2\right)-H\left(\theta-\frac{2\pi (n-1)}{N}-\phi_n/2\right)\right\}\nonumber\\
						&+\sum_{j=0}^{1}\left\{ H\left(\theta-2\pi j+\phi_1/2\right)-H\left(\theta-2\pi j-\phi_1/2\right)\right\},
\end{align}
where $H(\theta)$ is the Heaviside function.

The total wave field outside the slit cylinder is represented by $p_o$ which can be found as 
\begin{equation}
\label{ss_osol}
    p_o=H_0^{(1)}(k r)+\sum_{n=-\infty}^{+\infty}A_n H_n^{(1)}(k \hat{r})e^{i n \hat{\theta}},
\end{equation}
where $\hat{r}(r,\theta)$ and $\hat{\theta}(r,\theta)$ are the polar coordinates of receiver with origin placed at the centre of the scatterer and $A_n$ are unknown coefficients. The solution inside the slit cylinder is given by
\begin{equation}
\label{ss_isol}
    p_i=\sum_{n=-\infty}^{+\infty}\left[B_n J_n(k \hat{r}) + C_n Y_n(k \hat{r})\right]e^{i n \hat{\theta}},
\end{equation}
where coefficients $B_n$ and $C_n$ have to be found. The coefficient $C_n$ can be derived in terms of $B_n$ or set to zero according to the type of the core that results in the following 
\begin{align}
\label{ss_Cn}
	C_n&=B_n\got{C}_n,\; n\in\mathbb{Z},
\end{align}
where
\begin{align}
\label{ss_Cnforemp}
	\got{C}_n&=0,&\textrm{N-slit cylinder alone},\\
\label{ss_Cnforrcore}
	\got{C}_n&=-\frac{J'_n(k a_1)}{Y'_n(k a_1)},&\textrm{rigid core},\\
\label{ss_Cnforecore}	
	\got{C}_n&=-\frac{[J'_n(k R)]^2[1-(k_s R)^2+n^2]}{J'_n(k R) Y'_n(k R)[1-(k_s R)^2+n^2]+
    																																			[n^2-(k_s R)^2]\rho(\rho_s \pi R h)^{-1}},&\textrm{elastic shell core},
\end{align}
where primes (') denote the derivative with respect to polar coordinate $\hat{r}$. In equation \eqref{ss_Cnforecore} the elastic shell is described by frequency parameter $k_s=\omega/c_s$, density $\rho_s$, dilatational wave speed $c_s$, radius of the elastic shell mid-surface $R=(a_1 + a_2)/2$ and its half thickness $h_s=(a_1 - a_2)/2$. Here the dilatational wave speed $c_s$ for a thin elastic plate is given by
\begin{equation}
\label{am_dspped}
	c_s=\sqrt{\frac{E}{\rho_s\left(1-\nu^2\right)}},
\end{equation}
with Young's modulus $E$ and the Poison's ratio $\nu$. In this paper we consider thin viscoelastic shells made of latex \cite{SUOU}. For the numerical predictions the size of each elastic shell is always defined by the outer radius $a_1=0.02$ m and thickness $2 h_s=0.00025$ m.

In order to find the unknown coefficients $A_n$ we apply Graf's addition theorem \cite{Abramowitz} to the outer solution \eqref{ss_osol}. This enables us to express $(r,\theta)$ in terms of $(\hat{r},\hat{\theta})$ \cite{SUOU}. Then, substituting \eqref{ss_osol} and \eqref{ss_isol} into the boundary conditions \eqref{ss_BC} and taking the inner product $\displaystyle{\int_{0}^{2\pi}<\cdot>\exp(-i m \theta)d\theta}$ we arrive at infinite algebraic system of equations in $A_n,\,n\in\mathbb{Z},$ variables as follows 
\begin{align}
\label{ss_sys}
	&\sum_{n=-\infty}^{\infty}A_n\fbrl \delta_{m,n} 2 \pi h  {H_n^{(1)}}'(k r_o)\nonumber\\
							&\quad\quad- F_{n-m}\left[H_n^{(1)}(k r_o)-{H_n^{(1)}}'(k r_o) I_n\right]
							+\frac{k^2h}{2\pi} H_n^{(1)}(k r_o)\sum_{j=-\infty}^{\infty}F_{j-m}F_{n-j}I_j \fbrr=\\
	&\sum_{n=-\infty}^{\infty}H_n^{(1)}(k Q) e^{-i n (\pi+\alpha)}\fbrl-\delta_{m,n} 2\pi h J'_n(k r_o) \nonumber\\
							&\quad\quad+F_{n-m}\left[J_n(k r_o) - J'_n(k r_o)I_n\right] 
																						- \frac{k^2h}{2\pi} J_n(k r_o) \sum_{j=-\infty}^{\infty}F_{j-m}F_{n-j}I_j\fbrr,\;m\in\mathbb{Z},\nonumber
\end{align}
where $\delta_{m,n}$ is Kronecker delta, vector $\vc{Q} = Q (\cos\alpha,\sin\alpha)$ is the radius vector to the centre of the slit cylinder,
\begin{equation}
\label{ss_coefsI}
	I_n	= \frac{J_n(k r_i)+\got{C}_n Y_n(k r_i)}{J'_n(k r_i)+\got{C}_n Y'_n(k r_i)},\\
\end{equation}
and $F_n$ is the Fourier transform of function $f(\theta)$ given by
\begin{align}
\label{ss_Fprs}
	F_n=\left\{\begin{array}{ll}
							\displaystyle{\sum_{l=1}^N} \phi_l & \textrm{for}\quad n=0,\\
							\displaystyle{\frac{2}{n}\sum_{l=0}^{N-1}} \sin\left(\frac{n\phi_l}{2}\right) e^{-2 i n\pi l/N}& \textrm{for}\quad n\neq 0.
						\end{array}\right.
\end{align}
Note that factor $\got{C}_n$ and the geometrical parameters of the slits only appear in $I_n$ and $F_n$, respectively. This makes form of the system \eqref{ss_sys} invariant with respect to scatterer core type and the arrangement and size of the slits.

Taking $\phi_l=0,\,l\in\mathbb{Z},$ the solutions of system \eqref{ss_sys} is reduced to the case of scattering by rigid cylinder that is
\begin{equation}
\label{ss_coefAnRC}
	A_m=-\frac{J'_m(k r_o)}{{H_m^{(1)}}'(k r_o)}H_m^{(1)}(k Q) e^{-i m (\pi+\alpha)},\;m\in\mathbb{Z}.
\end{equation}

To find the numerical solution the infinite system \eqref{ss_sys} has to be truncated at some number $m,n,j=-M..M$. In general the convergence of the numerical solution is dependent on the radius $r_o$ of the N-slit cylinder,  number of slits and their angles $\phi_l$ (i.e. their length) as well as on the frequency. Reducing the angle $\phi_n$ of the slit results in faster convergence so that numerical solution approaches the value defined in \eqref{ss_coefAnRC}. It is found that for $k r_o < 5$ and $\phi_l<0.1,\,l\in\mathbb{Z},$ truncation number $M$ between 30 and 40 gives accurate results to less than three significant figures. Throughout this paper we use $M=35$.
 
\subsection{Results}\label{ssresults}

\begin{figure}[t]
		\center
		\subfigure[]{\includegraphics[scale=.7]{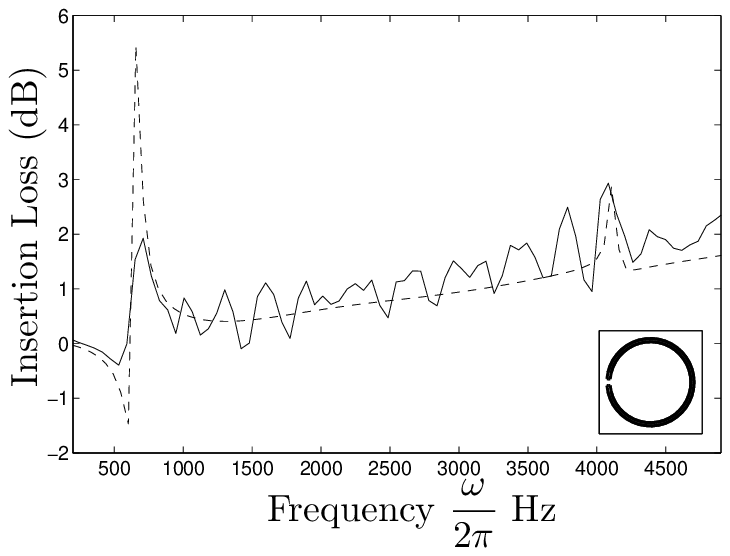}}\hspace{5mm}
		\subfigure[]{\includegraphics[scale=.7]{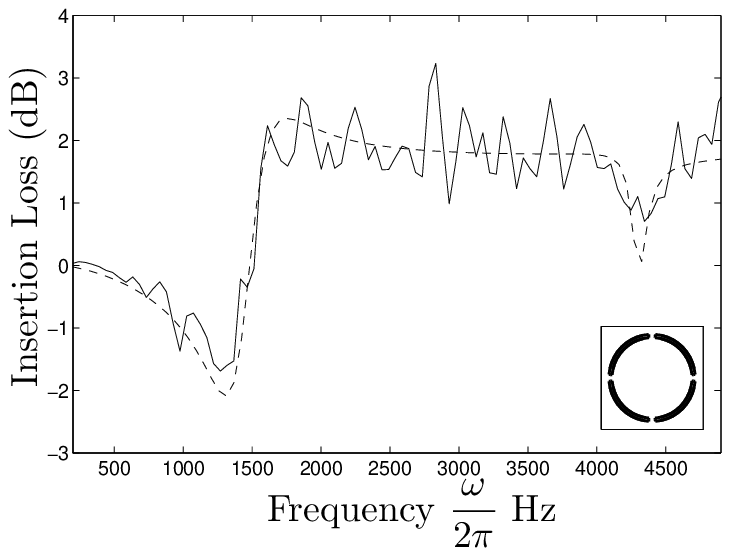}}\\
		\subfigure[]{\includegraphics[scale=.7]{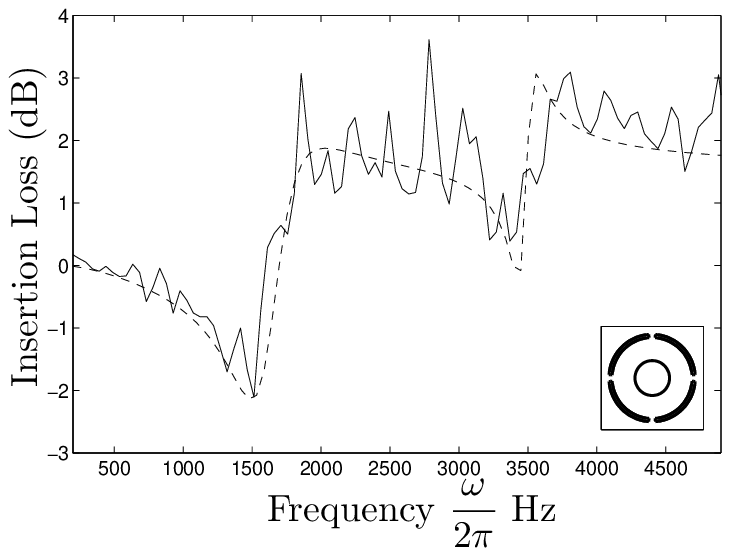}}\hspace{5mm}
		\subfigure[]{\includegraphics[scale=.7]{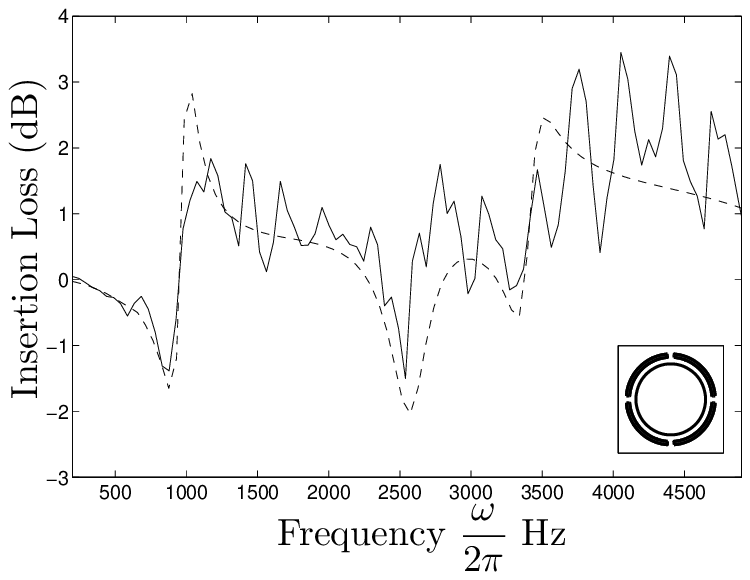}}
		
		\caption{Semi-analytical predictions (-~-~-~-) compared with data (--------) for single scatterers. Insertion losses are computed at the point $(1.5775,0)$ m. The rigid shell has outer radius $r_o=0.0275$ m, thickness $h=0.002$ m and $N$ identical slits of width $d_n\approx0.004\,{\rm m},\,n=1..N,$. (a) Single slit rigid cylinder (1S); (b) N-slit rigid cylinder with N=4 uniformly distributed slits (4S); (c) Concentric arrangement with outer 4S and inner rigid cylinders with $a_1=0.011{\rm m}$; (d) Concentric arrangement with outer 4S rigid cylinder and inner latex elastic shell.}
		\label{fig:ss_PVC_LTXs}
\end{figure}

Knowing the coefficients $A_n,\;n\in\mathbb{Z},$ enables us to compute scatterer insertion loss given by
\begin{equation}
\label{ss_IL}
	IL=20\log_{10}\left|\frac{H_0^{(1)}(k_o r)}{p_o}\right|,
\end{equation}
with acoustic potential $p_o$ found by formula \eqref{ss_osol}.

The insertion loss computed for a single N-slit rigid scatterer is compared with data in Figure~\ref{fig:ss_PVC_LTXs} (the experimental setup is described in section~\ref{experiment}). The results illustrated in Figures~\ref{fig:ss_PVC_LTXs}(a),(b) and (c) are obtained for N-slit cylinders alone and with a rigid core. For the given scatterer the prediction exhibits Helmholtz resonance in the low-frequency range (i.e. $k r_o < 1$). In both semi-analytical results and data, this resonant behavior is observed around $600$ Hz for a single slit rigid cylinder, see Figure~\ref{fig:ss_PVC_LTXs}(a), and around $1500$ Hz for a 4-slit rigid cylinder, see Figure~\ref{fig:ss_PVC_LTXs}(b). It is observed that increasing the number of slits results in significant shift of this Helmholtz-type resonance toward higher frequencies. The resonance can also be shifted by reducing the cavity area. This effect is seen in Figure~\ref{fig:ss_PVC_LTXs}(c) where the presence of a concentric internal rigid cylinder results in a shift of the resonance to a higher frequency $f \approx 1700$ Hz. 

The second type of the resonant behaviour observed in Figure~\ref{fig:ss_PVC_LTXs} is related to the cut off frequency of the acoustic mode in the circular/annular cavity. For the single slit rigid cylinder this resonance is found around $4200$ Hz (Figure~\ref{fig:ss_PVC_LTXs}(a)). The 4-slit rigid cylinder (Figure~\ref{fig:ss_PVC_LTXs}(b)) exhibits this type of resonance at a slightly higher frequency $f \approx 4400$ Hz. For the annular cavity formed by the concentric 4-slit rigid cylinder and rigid internal cylinder one of the cut off frequencies can be found around $3500$ Hz (Figure~\ref{fig:ss_PVC_LTXs}(c)).

Figure ~\ref{fig:ss_PVC_LTXs}(d) illustrates predicted and measured insertion losses for a 4-slit rigid cylinder with an inner elastic shell. The presence of the inner elastic shell with radius $a_0=0.02$ m causes resonant behavior to appear around $1000$ Hz, $2700$ Hz and $3400$ Hz. The first two resonances can be related to the interaction of the axisymmetric resonance of the elastic shell and the Helmholtz resonance. The frequency of the axisymmetric resonance of the elastic shell, which is originally at about 1200 Hz, is  decreased to about 1000 Hz by the concentric arrangement. The second resonance at $f \approx 2700$ Hz is related to the Helmholtz resonance. This frequency is higher than that for the empty slit cylinder due to the effective reduction in the size of the cavity.

\subsection{Low-frequency approximations}\label{ssapprox}

\begin{figure}[t]
		\center
		\subfigure[]{\includegraphics[scale=.7]{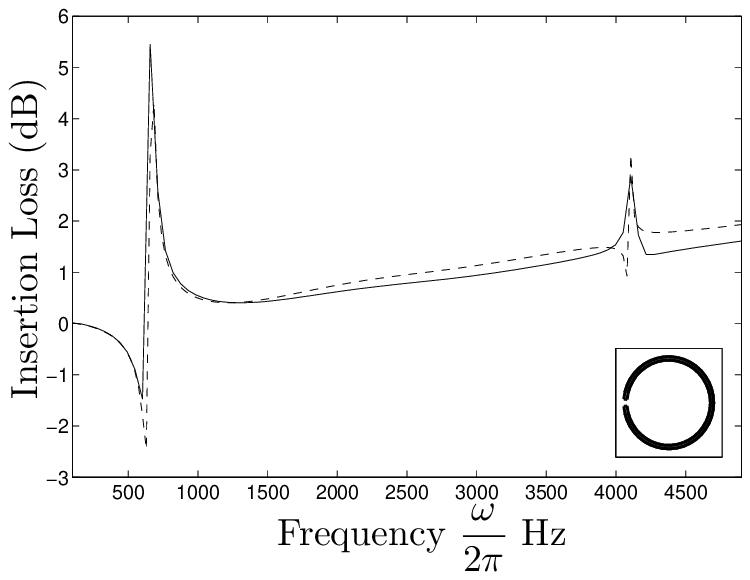}}\hspace{5mm}
		\subfigure[]{\includegraphics[scale=.7]{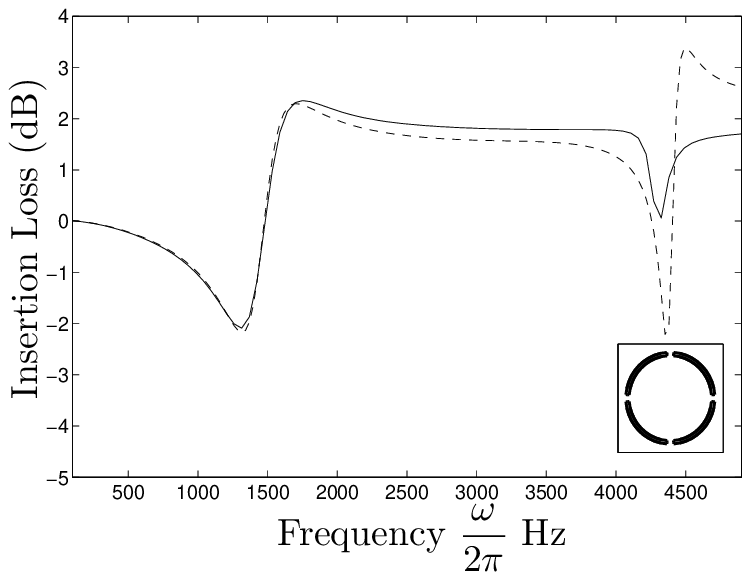}}\\
		\subfigure[]{\includegraphics[scale=.7]{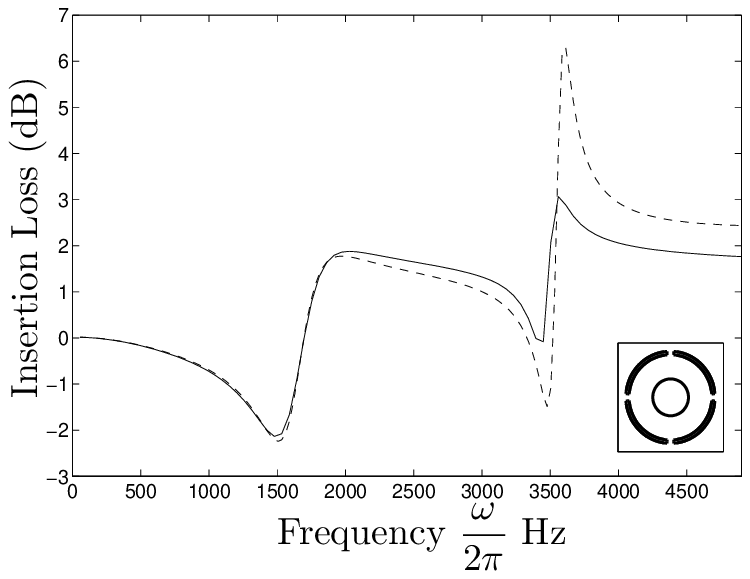}}\hspace{5mm}
		\subfigure[]{\includegraphics[scale=.7]{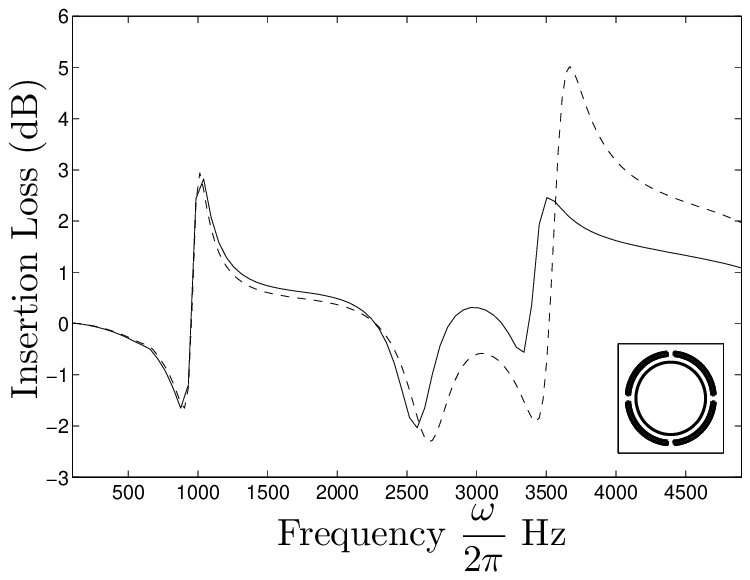}}
		
		\caption{Solution \eqref{ss_osol}-\eqref{ss_sys} (--------)  for single N-slit scatterer compared with its approximations (-~-~-~-) defined by \eqref{ss_apxc}, \eqref{ss_apxr} and \eqref{ss_apxAn}. Insertion loss is computed at the point $(1.5775,0){\rm m}$. (a) 1S; (b) 4S; (c) Concentric arrangement with outer 4S and inner rigid cylinders with $a_1=0.011{\rm m}$; (d) Concentric arrangement with outer 4S rigid cylinder and inner latex elastic shell.}
		\label{fig:ss_apxs}
\end{figure}

In the low-frequency range a simpler approximation can be used to describe scattering by N-slit cylinders. The approximation is based on the results derived by Horace Lamb \cite{Lamb} for the electromagnetic waves transmitted through a metallic grating. Here we only consider a case when N identical slits of width $d_n=d,\,n=1..N,$ are distributed periodically along the scatterer surface. In this case N-slit cylinder can be replaced by a homogeneous fluid shell. The latter is characterised by two effective parameters that are density $\rho_l$ and speed of sound $c_l$.

Similar approximations have been used for a periodic array of circular scatterers \cite{Hu}. The fluid layer is defined through the thickness $h_l=h \got{F}$ and density $\rho_l=\rho/\got{F}^2$ with filling fraction $\got{F}=N d/(2 \pi r_o)$ satisfying the following conditions
\begin{equation}
\label{ss_apxHu}
	\got{F} < \frac{1}{4} \quad\textrm{and}\quad \lambda=\frac c f > \got{F}.
\end{equation}
The parameters $h_l$ and $\rho_l$ were obtained by matching transmission coefficients derived for the infinite homogeneous and perforated plates. Having thickness as an effective parameter for the infinite plate gives quite accurate results in the low-frequency regime. However the fact that $h_l$ is not equal to the actual thickness of the element leads to unnecessary complications in the model for the cylindrical shell.

In contrast here the thickness of the equivalent fluid layer is made equal to the actual thickness of the N-slit cylinder and, therefore, the inner and outer radii of the corresponding fluid shell are uniquely defined through those of the N-slit cylinder. The obtained approximation gives first resonance position within $5\%$ of that for the N-slit cylinder provided that length of $d$ satisfies conditions \eqref{ss_apxHu}.

To find the unknown density $\rho_l$ and speed of sound $c_l$ we first write down the plane wave transmission coefficient of a fluid layer of thickness $h$ as
\begin{equation}
\label{ss_apxTfl}
	T=\frac{4 Z^{-1} e^{-i k h}}{(1+Z^{-1})^2 e^{-i k_l h} - (1-Z^{-1})^2 e^{i k_l h}},\quad Z=\frac{\rho_l c_l}{\rho c},
\end{equation}
where $k_l$ is the wave number in the fluid layer and $Z$ is its relative impedance. Note that coefficient $T$ is derived for the plane waves propagating perpendicular to the surface of the layer. In the low frequency regime when $k h \ll 1$ and $k_l h \ll 1$ the transmission coefficient can be approximated by
\begin{equation}
\label{ss_apxTflapx}
	T \approx \frac{Z^{-1}}{Z^{-1} - 0.5 i k_l h (1+Z^{-2})}.
\end{equation}

For a perforated plate such that $k d \ll 1$ the transmission coefficient $\hat{T}$ is described by oscillating piston-type waves propagating in slits. Using existing analytical results \cite{Shenderov} we arrive at
\begin{equation}
\label{ss_apxTsp}
	\hat{T}=\frac{4 \got{F} e^{-i k h}}{(1+\hat{Z}^{-1})^2 e^{-i k h} - (1-\hat{Z}^{-1})^2 e^{i k h}},
\end{equation}
with
\begin{equation}
\label{ss_apxZsp}
	\hat{Z}=\got{F}-\frac{i k d}{\got{F}^2 \pi^3}\sum_{n=1}^{\infty}\frac{\sin(\got{F} \pi n)}{n^3}=\got{F}-i k \Delta,
\end{equation}
where $\Delta$ is referred to as end correction. The lower limit of the end correction can be estimated from \cite{Lamb,Smits}
\begin{equation}
\label{ss_apxDelta}
	\Delta=\frac{d}{\pi} \log\frac{1}{\sin(\pi\got{F}/2)}.
\end{equation}
By assuming low frequency and small filling fraction (i.e. $\got{F} \ll 1$) the coefficient $\hat{T}$ is approximated as
\begin{equation}
\label{ss_apxTspapx}
	\hat{T} \approx \frac{\got{F}}{\got{F} - 0.5 i k h \left(1+\got{F}^2+2\Delta/h\right)}.
\end{equation}

In order to be able to replace the perforated plate by an effective fluid layer we equate $T$ and $\hat{T}$. Comparison of these two coefficients gives relative impedance $Z$ and the relationship between wave numbers $k_l$ and $k$ in the following forms
\begin{align}
\label{ss_apxZK}
	Z	 & = \frac{1}{\got{F}}\nonumber\\
	k_l &= \frac{k}{h}\left(h+\frac{2\Delta}{1+\got{F}^2}\right) \approx \frac{k}{h}\left(h + 2 \Delta\right).
\end{align}
It is then possible to derive the density and sound speed of the effective fluid layer as
\begin{align}
\label{ss_apxc}
	c_l &= \frac{h}{h+2\Delta} c\\
\label{ss_apxr}	
	\rho_l &= \frac{h+2\Delta}{h\got{F}} \rho.
\end{align}

To derive the solution of the appropriate scattering problem for the layered cylinder continuity boundary conditions have to be imposed at the faces $r=r_o, r_i$ of the fluid layer in the following form
\begin{align}
\label{ss_apxBC}
	p_l &= -\frac{\rho}{\rho_l} p_\alpha \nonumber\\
	\frac{\pd p_l}{\pd r} &=\frac{\pd p_\alpha}{\pd r},
\end{align}
where solutions $p_\alpha$ are given by \eqref{ss_osol} and \eqref{ss_isol} whereas solution inside the fluid layer is defined through the new set of constants $D_n$ and $E_n$, $n\in\mathbb{Z}$, as
\begin{equation}
\label{ss_apxlsol}
	p_l = \sum_{n=-\infty}^{+\infty}\left[D_n J_n(k_l \hat{r}) + E_n Y_n(k_l \hat{r})\right]e^{i n \hat{\theta}}.
\end{equation}
Note that conditions \eqref{ss_apxBC} and solution \eqref{ss_apxlsol} replace the boundary conditions \eqref{ss_BC} imposed at the faces of the N-slit cylinder.

Using boundary conditions \eqref{ss_apxBC} the unknown coefficients $A_n$, $D_n$,$E_n$ and $C_n$ can be found. The coefficient $A_n$ takes the following form
\begin{equation}
\label{ss_apxAn}
	A_n = -	Z_n H_n^{(1)}(k_o Q)e^{-i n \left(\pi+\alpha\right)},
\end{equation}
where
\begin{equation}
\label{ss_apxZn}
	Z_n = \frac{\rho_l W_1 W_3 J_n'(k r_o) - \rho J_n(k r_o) \left[W_1 J_n'(k_l r_o)+W_2 Y_n'(k_l r_o)\right]}
     				 {\rho_l W_1 W_3 \got{H}'_n(r_o) - \rho \got{H}_n(r_o) \left[W_1 J_n'(k_l r_o) + W_2 Y_n'(k_l r_o)\right]},
\end{equation}
within which
\begin{align}
\label{apx_W123}
	\got{H}_n(r_o)&= H_n^{(1)}(k r_o),\nonumber\\[8pt]
	W_1&= \left[-\rho Y_n'(k_l r_i) Y_n(k r_i) + \rho_l Y_n(k_l r_i) Y_n'(k r_i)\right]\got{C}_n
                            + \rho_l J_n'(k r_i)Y_n(k_l r_i) - \rho J_n(k r_i) Y_n'(k_l r_i),\nonumber\\[8pt]
	W_2&= \left[\rho J_n'(k_l r_i) Y_n(k r_i) - \rho_l J_n(k_l r_i) Y_n'(k r_i)\right]\got{C}_n
                            -\rho_l J_n'(k r_i) J_n(k_l r_i) + \rho J_n(k r_i) J_n'(k_l r_i),\nonumber\\[8pt]
	W_3&= \frac{W_1 J_n(k_l r_o) + W_2 Y_n(k_l r_o)}{W_1},
\end{align}
with $\got{C}_n$ defined in \eqref{ss_Cnforemp}.

To validate the low-frequency approximation its predictions are compared with those of the model presented in section~\ref{ssresults}.
In Figure \ref{fig:ss_apxs} the results are shown for the same types of scatterers. The accuracy of the approximation deteriorates with increase in frequency. However in the frequency range that is inside of the limit \eqref{ss_apxHu} the approximation reproduces the shift of the resonances observed for the concentric arrangement in Figures \ref{fig:ss_apxs}(c) and (d).

The low-frequency resonance displayed in Figure \ref{fig:ss_apxs} below 2000 Hz can be found by constructing approximation of the appropriate eigenvalue solution. In the eigenvalue problem the outer acoustic potential \eqref{ss_osol} is defined through Bessel functions of second kind only. The solution of this problem corresponds to zeros of the determinant of coefficient $A_n$ in \eqref{ss_apxAn} with $\got{H}_n(r_o)=Y_n(k r_o)$. To obtain the approximations of the resonant frequencies we need to consider two different cases defined by the core type factor $\got{C}_n$ in \eqref{ss_Cnforemp}-\eqref{ss_Cnforecore}. For N-slit cylinders that are either empty or concentric with an inner rigid cylinder the leading order approximations can be derived by assuming that
\begin{align}
\label{ss_apxHRpars}
	k r_o &= \got{O}(\beta),\;
	k_l r_o = \got{O}(\beta),\nonumber\\
	\frac{\rho}{\rho_l} &= \got{O}(\beta^2),\nonumber\\	
\end{align}
where $\beta \ll 1$ is a small parameter. These four relations define the frequency regime and contrast between fluid layer and outer/inner environment. Note that according to equation \eqref{ss_apxr} and the assumption that $\got{F}\ll 1$, density $\rho_l$ is always much greater than $\rho$.

Expanding Bessel functions and collecting similar orders of smallness the first resonant frequency i.e. $n=0$ can be approximated by
\begin{equation}
\label{ss_apxHRres}
	f = \frac{c}{2\pi} \sqrt{\frac{2\rho}{\rho_l (r_i^2-a_1^2)\log(r_o/r_i)}}.
\end{equation}
The resulting values of resonant frequency can be further approximated by using the fact that ratios $h/r_i$ and $h/r_o$ are small. This follows from the assumption made in the beginning of section \ref{ssformulation}. The expansion $\log(r_o/r_i) = h/r_i + \got{O}(h^2/r_i^2)$ and substitution of the equation \eqref{ss_apxr} give
\begin{equation}
\label{ss_apxHRresapx}
	f \approx \frac{c}{2\pi} \sqrt{\frac{N d}{\pi (r_i^2-a_1^2) (h + 2\Delta)}}.
\end{equation}
For a single slit $N=1$ equation \eqref{ss_apxHRresapx} reduces to the well-known value of Helmholtz resonance. From equation \eqref{ss_apxHRresapx} it is also seen that increasing the number of slits leads to the shift of the Helmholtz resonance toward higher frequencies by approximately $\sqrt{N}$. This is in accordance with the results observed in Figures~\ref{fig:ss_apxs} (a) and (b) which show a shift of Helmholtz resonance frequency by a factor of 2 as N increases from 1 to 4.

Next is considered the case of the concentric N-slit cylinder and elastic shell which is described by factor $\got{C}_n$ in equation \eqref{ss_Cnforecore}. The interaction of the axisymmetric and Helmholtz resonances leads to the considerable shift of the low-frequency resonance as well as the high-frequency resonances. The new position of the low-frequency resonance can be approximated by introducing the following order of smallness of the physical and geometrical parameters 
\begin{align}
\label{ss_apxSpars}
	k r_o &= \got{O}(\beta), \;	k_l r_o = \got{O}(\beta),\nonumber\\	
	\frac{\rho}{\rho_s} &= \got{O}(\beta^3), \; \frac{\rho_l}{\rho_s} = \got{O}(\beta),\nonumber\\
	\frac{c_s}{c} &= \got{O}(\beta), \; \frac{c_s}{c_l} = \got{O}(\beta),\nonumber\\
	\frac{h_s}{R} &= \got{O}(\beta),
\end{align}

For $n=0$ the leading order approximation of the resonant frequency becomes
\begin{align}
\label{ss_apxSres}
	f = &\frac{c}{2\pi} \fbrl \frac{1}{2 h_s R (r_i^2-R^2) \log(r_o/r_i)} \sbrl
				\left(\frac{\rho}{\rho_s} r_i^2 + \frac{c_s^2}{c^2} \frac{h_s}{R} (r_i^2-R^2)\right)\log\frac{r_o}{r_i} + 2 \frac{\rho}{\rho_l} h_s R\nonumber\\
			&\quad\quad\pm \sqrt{\left[\left(\frac{\rho}{\rho_s}r_i^2 + \frac{c_s^2}{c^2} \frac{h_s}{R} (r_i^2-R^2)\right)\log\frac{r_o}{r_i} 
																																														- 2 \frac{\rho}{\rho_l} h_s R\right]^2
							+ 8 \frac{\rho}{\rho_l}\frac{\rho}{\rho_s} h_s R^3 \log\frac{r_o}{r_i}} \sbrr \fbrr^{1/2}.
\end{align}
In this equation the presence of two solutions explains the interaction of the first two low-frequency resonances. Two limiting cases can be constructed which show the degeneration of the two resonances into single one. It can be seen that if radii $r_i$ and $r_o$ tend to infinity the equation \eqref{ss_apxSres} transforms to the resonance of the elastic shell under membrane compression loading \cite{SUOU} i.e.
\begin{align}
\label{ss_apxSonlyres}
	f = \frac{c}{2\pi} \frac{1}{R} \sqrt{\frac{c_s^2}{c^2} + \frac{R}{h_s}\frac{\rho}{\rho_s}}.
\end{align}
Comparison of the resonant frequency predicted by \eqref{ss_apxSonlyres} for an elastic shell made of latex with the smallest value in \eqref{ss_apxSres} obtained for the concentric 4S rigid cylinder and latex elastic shell enables us to estimate the shift of the resonance observed in the composite. The numerical value of the axisymmetric resonance \eqref{ss_apxSonlyres} is $f \approx 1270$ Hz whereas the lowest axisymmetric resonance of the composite is $f \approx 1060$ Hz. Thus a reduction of the frequency of the axisymmetric elastic shell resonance by about 200 Hz is predicted.


Equation \eqref{ss_apxSres} can also be reduced to the form of the resonant frequency \eqref{ss_apxHRres} by assuming higher contrast in \eqref{ss_apxSpars} between the fluid and elastic media.

\section{Arrays of scatterers}\label{aos}

\subsection{Formulation}

The method developed in the previous section can now be adapted to finite arrays of scatterers. We first formulate the method of multiple scattering that is described by the superposition of scattered wave fields of each element of the array \cite{Umnova} and boundary conditions \eqref{ss_BC} imposed on the surface of all scatterers in the array. As a result the solution $p_o$ of the Helmholtz equation \eqref{ss_Helmholtz} in the acoustic environment outside scatterers takes the following form
\begin{equation}
\label{aos_tfield}
	p_o(r,\theta)=H_0^{(1)}(k r) + \sum_{m=1}^{\got{N}}\sum_{n=-\infty}^{+\infty}A_n^m H_n^{(1)}(k \hat{r}_m)\exp(i n \hat{\theta}_m),
\end{equation}
where $\got{N}$ is the number of scatterers in the array, variables $\hat{r}_m(r,\theta)$, $\hat{\theta}_m(r,\theta)$ are the polar coordinates with origin in the centre of scatterer of index $m$ and $A_n^m$ are unknown coefficients. The form of the solution inside of the m-th slit cylinder is the same as in the case of single scatterer problem that is equation \eqref{ss_isol}.

Using Graf's addition theorem we can rewrite solution \eqref{aos_tfield} in terms of polar coordinates $(\hat{r}_m,\hat{\theta}_m)$ of m-th scatterer that gives
\begin{align}
\label{aos_Gtfield}
	p_o(\hat{r}_m,\hat{\theta}_m)&=\sum_{n=-\infty}^{+\infty}\fbrl J_n(k \hat{r}_m)H_n^{(1)}(k Q_m)e^{-i n \left(\pi+\alpha_m\right)}
			 + A_n^m H_n^{(1)}(k \hat{r}_m)\nonumber\\
			&+ \sum_{p=1,\,p \neq m}^{\got{N}}\sum_{q=-\infty}^{+\infty}A_q^p J_n(k \hat{r}_m) H_{q-n}^{(1)}(k Q_{mp})e^{i (q-n) (\alpha_{mp}+\pi)}\fbrr
																	e^{i n \hat{\theta}_m},\;m=1..\got{N}
\end{align}
where vector $\vc{Q_m}=Q_m(\cos\alpha_m,\sin\alpha_m)$ is the radius vector to the centre of m-th scatterer, vector $\vc{Q_{mp}}=Q_{mp}(\cos\alpha_{mp},\sin\alpha_{mp})$ defines the position of p-th scatterer with respect to m-th scatterer and $Q_{m},\, Q_{mp}>\hat{r}_m$ that is the requirement of addition theorem. Expansion \eqref{aos_Gtfield} along with the inner solution \eqref{ss_isol} are subject to the boundary conditions \eqref{ss_BC}. This results in the infinite algebraic system of equations in $A_n^m,\,m=1..\got{N},\,n\in\mathbb{Z},$ variables given by
\begin{align}
\label{aos_sys}
	&\sum_{n=-\infty}^{\infty} A_n^p \fbrl \delta_{m,n} 2 \pi h  {H_n^{(1)}}'(k r_{o,p})\nonumber\\
							&\quad\quad\quad- F_{n-m}^p\left[H_n^{(1)}(k r_{o,p})-{H_n^{(1)}}'(k r_{o,p}) I_n^p\right]
							+\frac{k^2h}{2\pi} H_n^{(1)}(k r_{o,p})\sum_{j=-\infty}^{\infty}F_{j-m}^p F_{n-j}^p I_j^p \fbrr\nonumber\\
	&+\sum_{s=1,\,s \neq p}^{\got{N}} \sum_{n=-\infty}^{\infty} \sum_{v=-\infty}^{\infty} 
							A_n^s H_{n-v}^{(1)}(k Q_{ps}) e^{-i (n-v) (\pi+\alpha_{ps})}\fbrl \delta_{m,v} 2 \pi h  J'_n(k r_{o,p}) \nonumber\\
							&\quad\quad\quad- F_{v-m}^p \left[J_n(k r_{o,p})-J'_n(k r_{o,p}) I_n^p\right]
							+\frac{k^2h}{2\pi} J_v(k r_{o,p})\sum_{j=-\infty}^{\infty}F_{j-m}^p F_{v-j}^p I_j^p \fbrr=\\
	&\sum_{n=-\infty}^{\infty}H_n^{(1)}(k Q_p) e^{-i n (\pi+\alpha_p)}\fbrl-\delta_{m,n} 2\pi h J'_n(k r_{o,p}) \nonumber\\
							&\quad\quad\quad+F_{n-m}\left[J_n(k r_{o,p}) - J'_n(k r_{o,p})I_n^p\right] 
																						- \frac{k^2h}{2\pi} J_n(k r_{o,p})
																						 \sum_{j=-\infty}^{\infty}F_{j-m}^p F_{n-j}^p I_j^p\fbrr,\;m\in\mathbb{Z},\;p=1..\got{N},\nonumber
\end{align}
where factors $I_m^p$ and $F_m^p$ are given by equations \eqref{ss_coefsI} and \eqref{ss_Fprs}, respectively. The superscript $p$ relates these factors to p-th scatterer. To find unknown coefficients we again use truncated system of $\got{N}(2M+1)$ algebraic equations. Note that for the slit angles $\phi_l^p=0,\,l\in\mathbb{Z},\,p=1..\got{N},$ equations \eqref{aos_sys} are transformed to the well-known algebraic system describing multiple scattering problem for an array of rigid cylinders \cite{Umnova}.

\subsection{Results}\label{aosresults}

\begin{figure}
		\center
		\subfigure[]{\includegraphics[scale=.7]{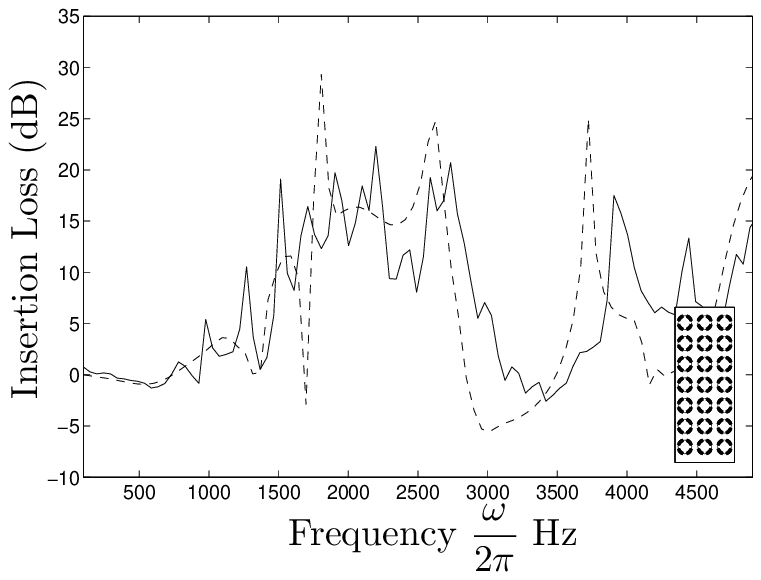}}\hspace{5mm}
		\subfigure[]{\includegraphics[scale=.7]{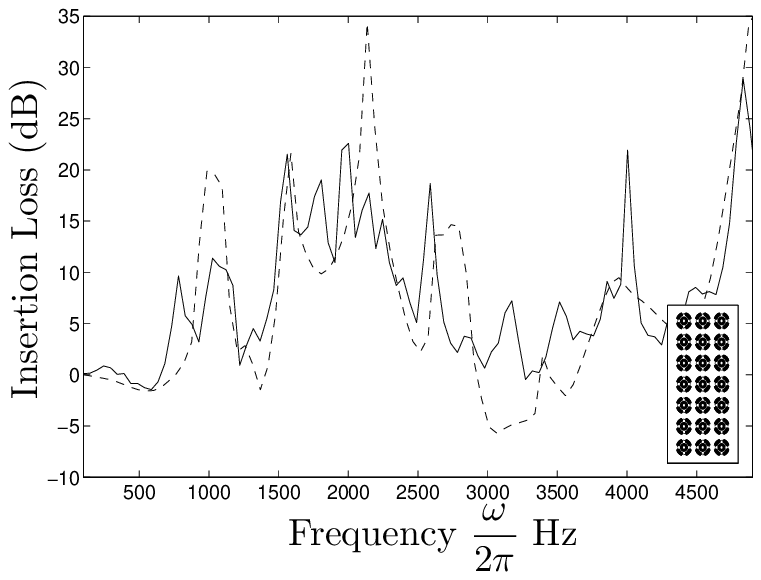}}
		
		\caption{Semi-analytical predictions (-~-~-~-) compared with data (--------) for $7 \times 3$ arrays of scatterers. Distances from array to the source and receiver are 1.5 m and 0.05 m, respectively. The scatterers are arranged in square lattice with lattice constant $L=0.086$ m. (a) Empty 4-slit rigid cylinders (4S); (b) Concentric outer 4S cylinder and inner latex elastic shell.}
		\label{fig:aos_PVC_LTXs}
\end{figure}

Solving the algebraic system of equations \eqref{aos_sys} enables us to compute the total wave field \eqref{aos_tfield} and insertion loss \eqref{ss_IL}. The latter is analysed in this section.

Figure~\ref{fig:aos_PVC_LTXs}(a) shows measured and predicted insertion loss spectra for a $7\times3$ array of 4-slit cylinders. The band gap effect related to the cavity resonance (see Figure~\ref{fig:ss_PVC_LTXs}) and associated with a high insertion loss peak is predicted and observed in the vicinity of $f=1500$ Hz. This effect is followed by the insertion loss peak related to the first Bragg band gap which is observed around $f \approx 2000$ Hz.

In Figure~\ref{fig:aos_PVC_LTXs}(b) the insertion loss is computed for the array of composite scatterers consisting of concentric 4S rigid cylinders and latex elastic shells. The coupled resonances of this composite described in section \ref{ssresults} generate various band gaps that result in high insertion loss peaks. One of these peaks is observed around $1000$ Hz and is related to the band gap due to the shifted axisymmetric resonance of latex elastic shell. The insertion loss peak associated with the shifted Helmholtz resonance is observed around $2600$ Hz and it follows the first Bragg band gap peak observed around $f \approx 2000$ Hz.

\subsection{Low-frequency approximations}

\begin{figure}
		\center
		\subfigure[]{\includegraphics[scale=.7]{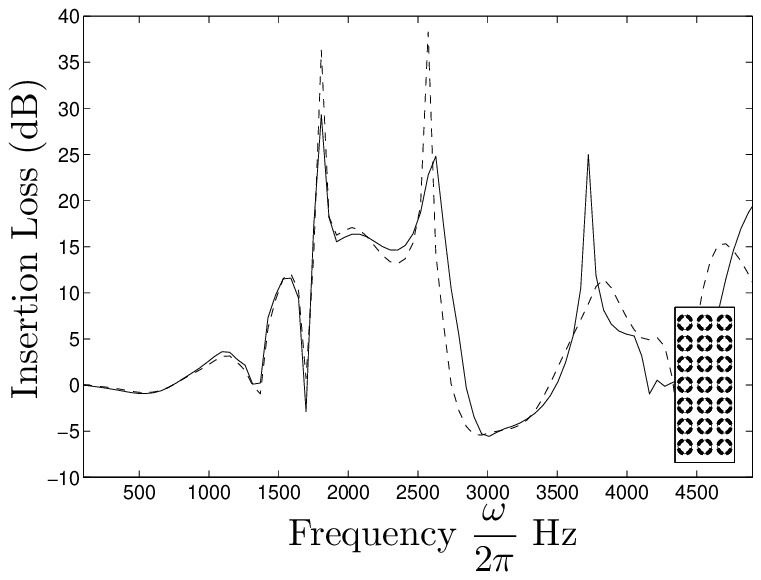}}\hspace{5mm}
		\subfigure[]{\includegraphics[scale=.7]{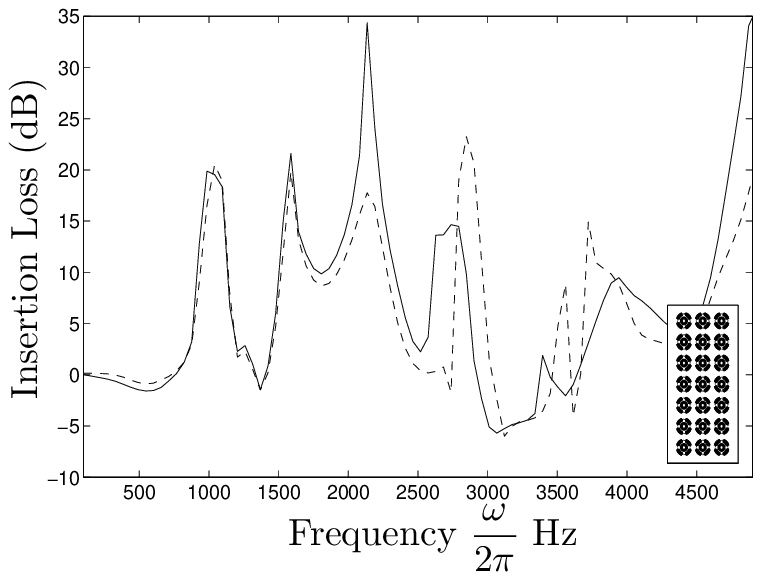}}
		
		\caption{Semi-analytical predictions for $7 \times 3$ array of composites (--------) compared with their low-frequency approximations (-~-~-~-). Distances from scatterer to the source and receiver are 1.5 m and 0.05 m, respectively. (a) 4S; (b) Concentric outer 4S rigid cylinder and inner latex elastic shell.}
		\label{fig:aos_apx_PVC_LTXs}
\end{figure}

\begin{figure}
		\center
		\subfigure[]{\includegraphics[scale=.7]{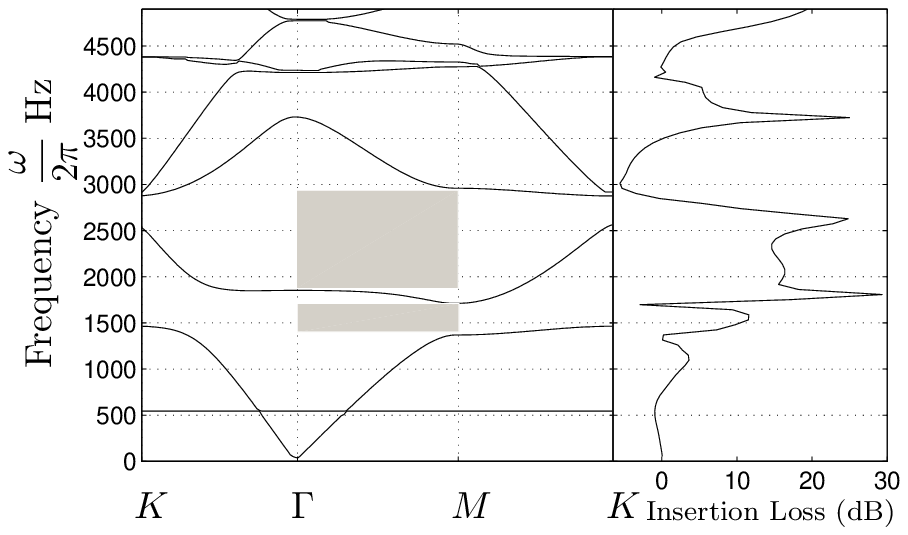}}
		\subfigure[]{\includegraphics[scale=.7]{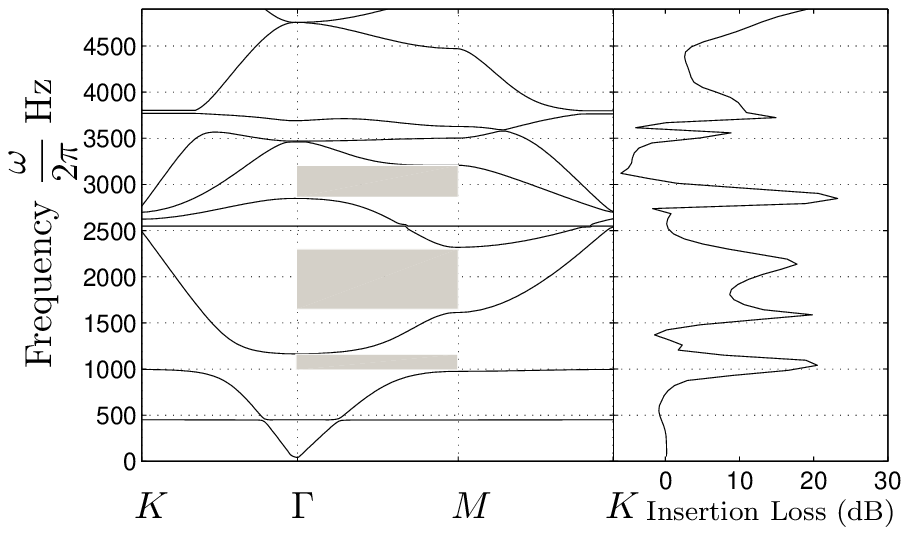}}
		\hspace{-5mm}
		
		\caption{Band diagram for the infinite periodic array of composites (left inset) compared with the low-frequency approximations of their finite $7 \times 3$ arrangement (right inset). All geometrical parameters are identical to those used in Figure \ref{fig:aos_apx_PVC_LTXs}. The enclosed shells in the infinite periodic structure are taken without viscous effects and their Young's modulus is identical to that used in \cite{SUOU}, Figure 6. (a) 4S; (b) Concentric outer 4S rigid cylinder and inner latex elastic shell.}
		\label{fig:aos_ps_PVC_LTXs}
\end{figure}

Here the model of equivalent fluid layer is applied to the problem of multiple scattering. Using the low-frequency approximations \eqref{ss_apxc} and \eqref{ss_apxr} together with the boundary conditions \eqref{ss_apxBC} one can derive the analogue of the algebraic system of equations \eqref{aos_sys} that is
\begin{equation}
\label{aos_apxsys}
	A_n^m+\sum_{p=1,\,p \neq m}^{\got{N}}\sum_{q=-\infty}^{\infty}A_q^p Z_q^p H_{q-n}^{(1)}(k_o Q_{mp})e^{i (q-n) (\pi+\alpha_{mp})}
																														=-H_n^{(1)}(k_o Q_m)e^{-i n \left(\pi+\alpha_m\right)},
\end{equation}
where $Z_q^p$ is given by \eqref{ss_apxZn} for the scatterer with index $p=1..\got{N}$ \cite{Umnova}. The truncated linear system of $\got{N}(2M+1)$ equations is solved to find the unknown coefficients $A_n^m$ and, as a result, the insertion loss \eqref{ss_IL}. The truncation number $M$ is taken between 5 and 7. This results in substantial reduction in the computational time compared to that for the full problem \eqref{aos_sys}.

In Figure \ref{fig:aos_apx_PVC_LTXs} insertion loss is computed for the same arrays as considered in section \ref{aosresults}. It is observed that replacement of the N-slit rigid cylinder by the equivalent fluid layer gives results accurate to within $5\%$ in the frequency interval $f\in(0,2000)$ Hz. Therefore, for the given geometry of the array the proposed approximations accurately predict the presence of the band gaps below and including the first Bragg band gap.

The band diagrams plotted in Figure \ref{fig:aos_ps_PVC_LTXs}(a) and (b) are aligned with the insertion loss spectra obtained previously in Figure \ref{fig:aos_apx_PVC_LTXs}. It is demonstrating that the band gaps coincide with the maxima of insertion loss. The diagrams also show that the first band gaps (at 1500 Hz in Figure \ref{fig:aos_ps_PVC_LTXs}(a) and at 1000 Hz in Figure \ref{fig:aos_ps_PVC_LTXs}(b)) are complete. This results in angular independent insertion loss peaks bounded in the frequency interval of the complete band gaps.

\section{Laboratory measurements}\label{experiment}

\subsection{Cylinder constructions}

\begin{figure}[ht]
		\center		
		\subfigure[]{\includegraphics[scale=.2]{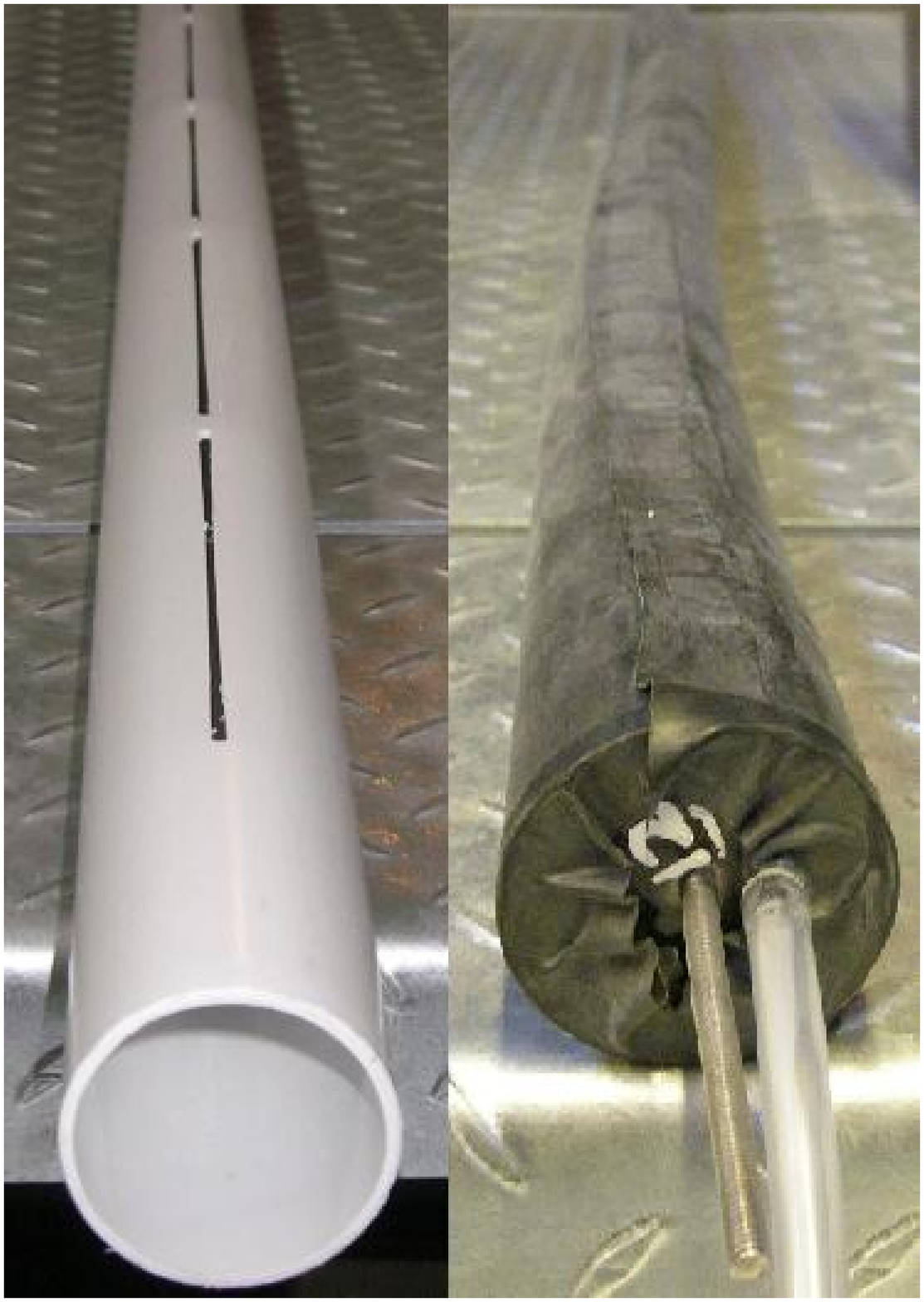}}
		\subfigure[]{\includegraphics[scale=.4]{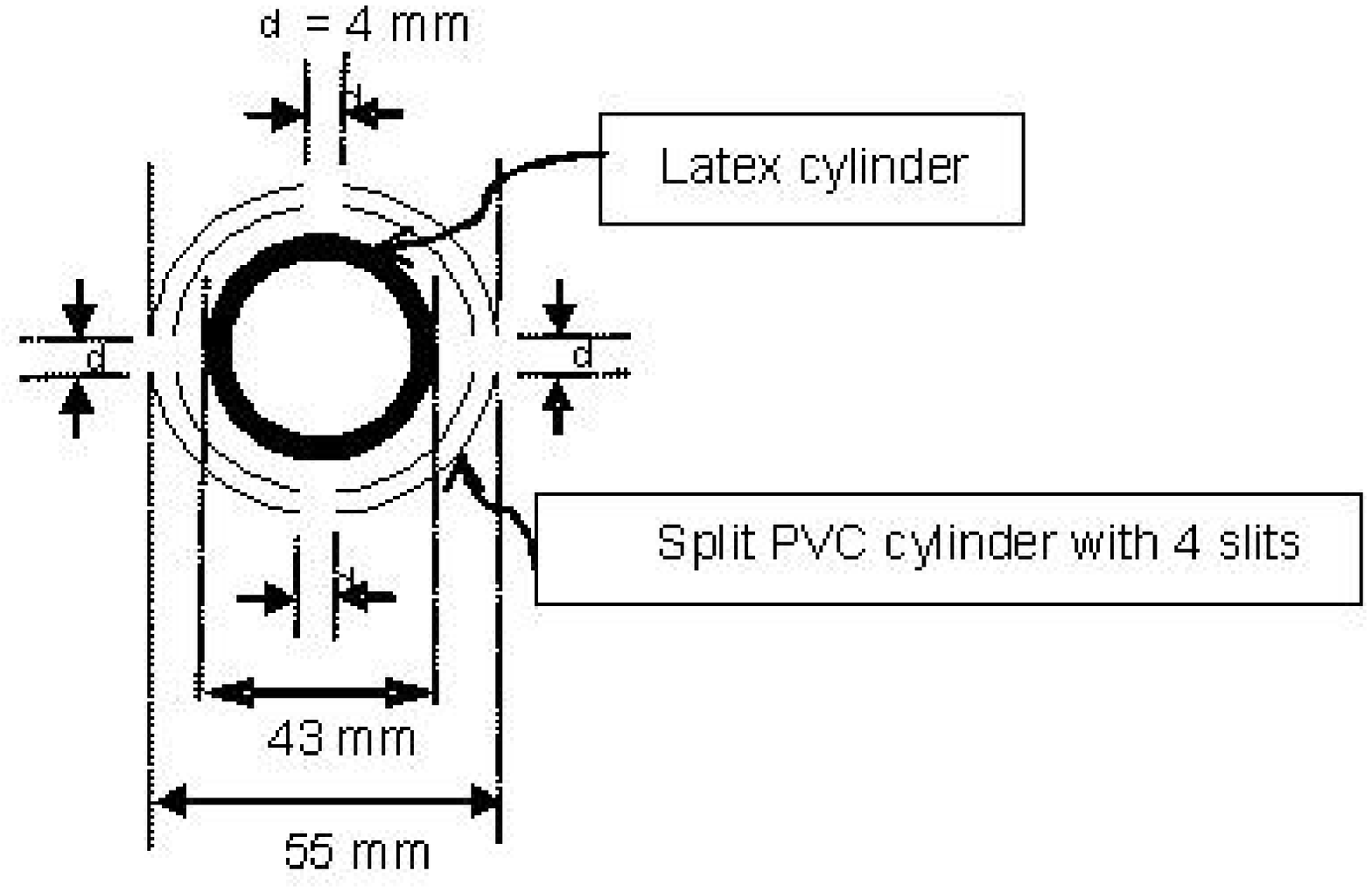}}		
		
		\caption{(a) Components of the resonating scatterer: (on the left) photograph of PVC pipe showing the discontinuous form of a (single) slit and (on the right) photograph of latex elastic shell. (b) Cross section of the concentric arrangement of inner latex cylinder and outer PVC cylinder with 4 symmetrically-placed slits }
		\label{fig:exp_ss}
\end{figure}

2 m long 0.25 mm thick Latex sheets have been formed into cylinders with outer diameter of 43 mm. This has been achieved by overlapping edges by a few mm and gluing them together. Slits have been gouged from the walls of 55 mm outer diameter PVC pipes. To ensure the structural integrity of the pipes, rather than making continuous slits along the complete length of the pipes, the slits were made in sections of approximately 20 cm length separated by about 2 cm (see Figure~\ref{fig:exp_ss}(a)). Concentric arrangements of pairs of latex and 4-slit PVC cylinders were formed as shown in Figure~\ref{fig:exp_ss}(b). To ensure the central location of each 27.5 mm outer diameter latex cylinder within the associated 4-slit PVC pipe, each end of the latex cylinder was secured on to a Perspex cap which fitted inside the PVC pipe. Smaller PVC pipes with inner diameters of 22 mm were used inside the 4-slit pipes for some measurements.

\subsection{Measurement system and data analysis}

\begin{figure}
		\center		
		\subfigure[]{\includegraphics[scale=.3]{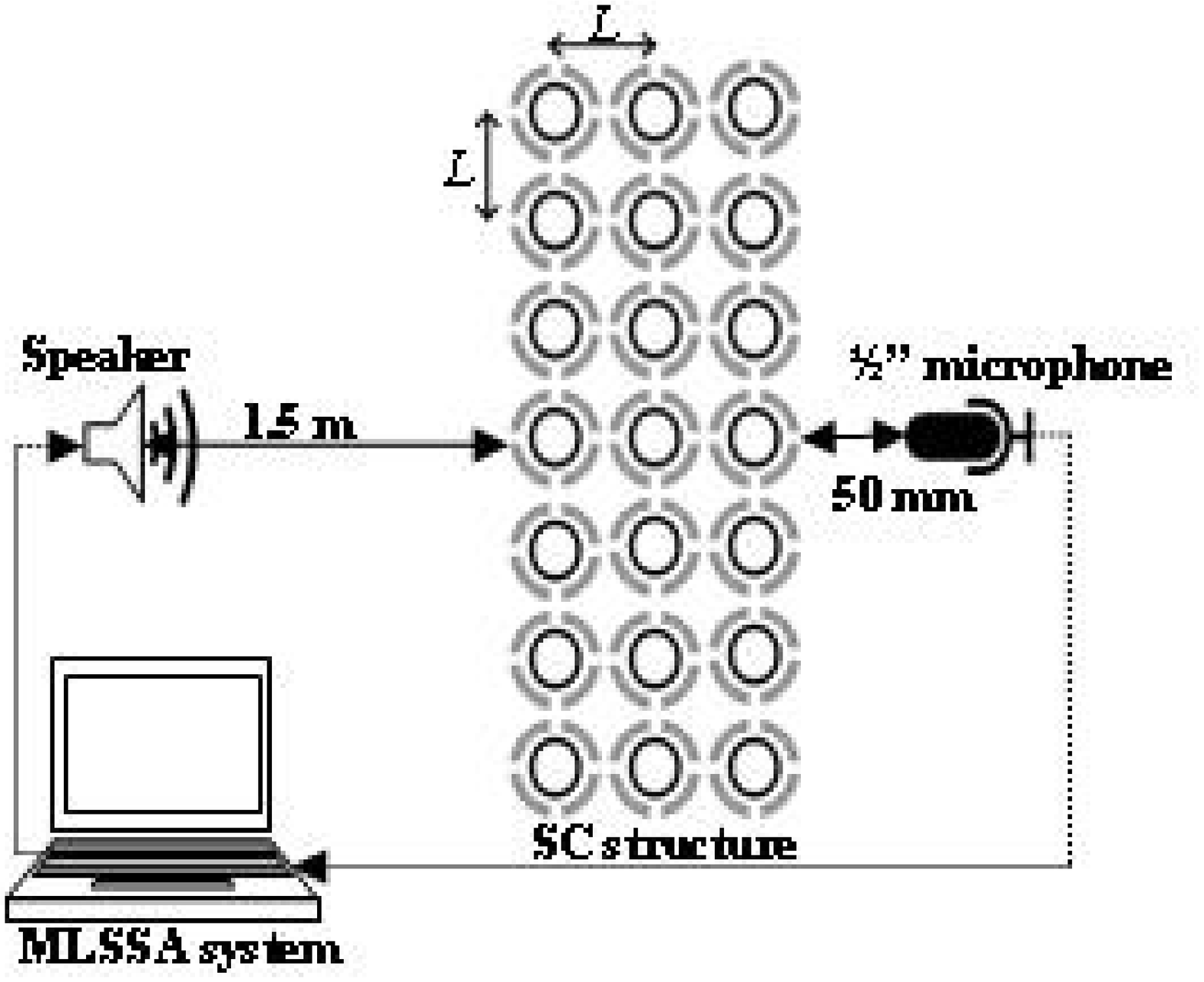}}
		\subfigure[]{\includegraphics[scale=.35]{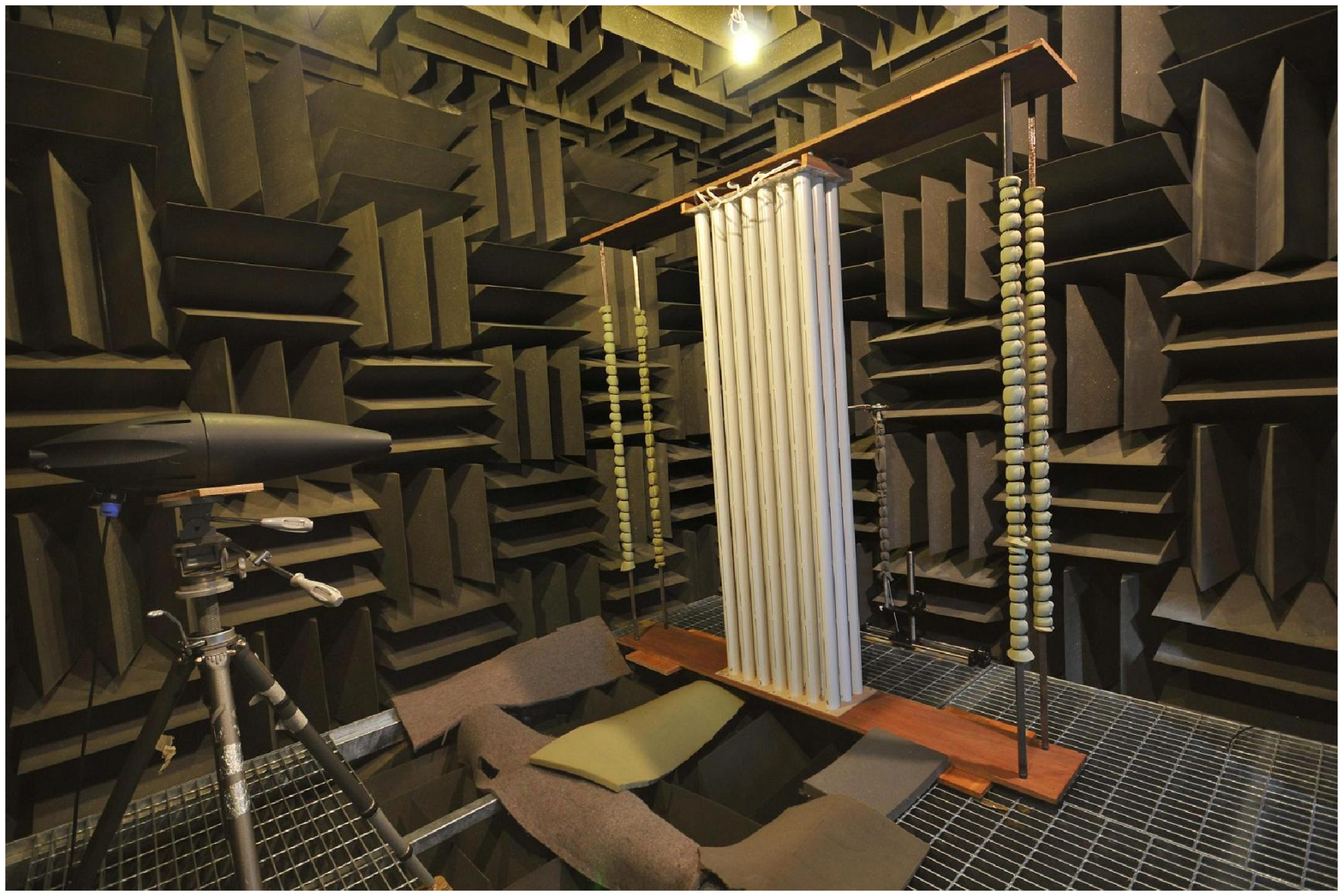}}		
		
		\caption{(a) plan of source, receiver and array (b) $7\times3$ array of concentric PVC pipe and latex cylinders in the anechoic chamber.}
		\label{fig:exp_aos}
\end{figure}

The sound source was a Bruel $\&$ Kjaer point source loudspeaker controlled by a Maximum-Length Sequence System Analyzer (MLSSA) system enabling determination of impulse responses. Measurements were taken of the insertion loss (IL) spectra for single cylinders and arrays of cylinders in an anechoic chamber. Figures~\ref{fig:exp_aos}(a) and (b) show an example measurement arrangement. Supports for the 2 m long cylinders were provided by holed wooden boards at the top and base of each array. The lattice constant (L) for the arrays of 4-slit cylinders and concentric cylinders was 80 mm.

To maintain their shape and vertical orientation, the latex cylinders were slightly inflated above atmospheric pressure through a common pipe connecting to a small pump. During the array measurements, the receiver microphone was positioned 50 mm from the nearest face of the array but on the opposite side to the source. The loudspeaker was placed 1.5 m away from the array, such that the source-receiver axis was normal to the array orientation (see Figure~\ref{fig:exp_aos}(a)). Both source and receiver were 1.2 m above the floor of the chamber which had sections removed to reduce unwanted reflections (see Figure~\ref{fig:exp_aos}(b)). Insertion loss spectra were calculated by subtracting signals received without and with the cylinder array present but with the support structure in place on both occasions.

\section{Concluding remarks}

The acoustical properties of three forms of 2D resonators have been investigated: empty N-slit pipes, a concentric arrangement with rigid pipe inner and 4-slit pipe outer and a concentric arrangement of an elastic shell inner and 4-slit outer pipe. A theoretical formulation uses boundary conditions dependent on polar angle to represent arbitrary positioned slits. For simplicity, in the studied geometries the slits are positioned symmetrically. Jump boundary conditions imposed on the slit interface are used to represent the solution inside the slits. The versatility of the proposed method can be applied to various type of 2D resonators with concentric multilayered/solid cylinder inner.

It has been found theoretically and experimentally that increasing the number of slits in an empty pipe causes an increase in the frequency of the Helmhotz-type resonance. A low frequency approximation which models the slit cylinder by an equivalent fluid layer predicts that the frequency increase is proportional to the square root of the number of slits and this has been confirmed experimentally. The concentric arrangements result in resonances associated with both circular and annular cavities. With an inner elastic shell, an additional axisymmetric resonance of the shell is preserved but modified by the presence of the outer 4-slit pipe. Coupling between the components of the concentric arrangement results in shifts in the resonant frequencies corresponding to each element of the composite configuration. This is similar to the effect that is observed in mass-spring systems with multiple degrees of freedom. A low frequency approximation for the acoustical properties of the concentric arrangement with an elastic shell inner correctly predicts the observed frequency shift in the axisymmetric resonance.

When used in periodic arrays the concentric arrangements with inner elastic shells and outer 4-slit cylinders result in additional sound attenuation in the low-frequency range below the first Bragg band gap while still preserving the Bragg band gaps. This arrangement is more practical that the use of unprotected elastic shells so is potentially useful basis for a sonic crystal barrier design.

\section*{Acknowledgments}
This work was supported by the EPSRC research grants EP/E063136/1 and EP/E062806/1. Authors are grateful for this support.



\end{document}